# Superconductivity of powder-in-tube $Sr_{0.6}K_{0.4}Fe_2As_2$ wires

Yanpeng Qi, Xianping Zhang, Zhaoshun Gao, Zhiyu Zhang, Lei Wang, Dongliang Wang, Yanwei Ma[*]

Key Laboratory of Applied Superconductivity, Institute of Electrical Engineering, Chinese Academy of Sciences, P. O. Box 2703, Beijing 100190, China

**Abstract:**

Nb-sheathed $Sr_{0.6}K_{0.4}Fe_2As_2$ superconducting wires have been fabricated using the powder-in-tube (PIT) method for the first time and the superconducting properties of the wires have been investigated. The transition temperature ($T_c$) of the $Sr_{0.6}K_{0.4}Fe_2As_2$ wires is confirmed to be as high as 35.3 K. Most importantly, $Sr_{0.6}K_{0.4}Fe_2As_2$ wires exhibit a very weak $J_c$-field dependence behavior even the temperature is very close to $T_c$. The upper critical field $H_{c2}(0)$ value can exceed 140 T, surpassing those of $MgB_2$ and all the low temperature superconductors. Such high $H_{c2}$ and superior $J_c$-field performance make the 122 phase SrKFeAs wire conductors a powerful competitor potentially useful in very high field applications.

---

[*] Author to whom correspondence should be addressed; E-mail: ywma@mail.iee.ac.cn

The recent discovery of the iron-based superconductor REFeAsO$_{1-x}$F$_x$ (where RE=La, Ce, Pr, Nd, Sm, Gd) has triggered great interest in the scientific community.[1-8] Similar to the cuprates, the FeAs layer is thought to be responsible for superconductivity and RE-O layer is a carrier reservoir layer to provide electron carrier. Thus, a new class of REFeAsO$_{1-x}$F$_x$ materials (FeAs-1111 phase) with a layered structure and relatively high superconducting $T_c$ values (up to 55 K) was born, rivaling the doped superconducting cuprates. More recently, BaFe$_2$As$_2$ compound with a new structure (FeAs-122 phase) was discovered, for instance, the superconductivity with $T_c$=38 K was found in K-doped BaFe$_2$As$_2$,[9] then superconductivity was reported in SrFe$_2$As$_2$, CaFe$_2$As$_2$ and EuFe$_2$As$_2$ by appropriate substitution or under pressure.[10-15] The $A$Fe$_2$As$_2$ (where $A$ = Ba, Sr, Ca, Eu) compound is built up with identical FeAs layers separated by $A$ instead of RE-O layers, so there are double Fe-As layers in unit cell.

New iron-based layered superconductors have much high upper critical fields ($H_{c2}$),[16, 17] indicating a very encouraging application of new superconducting wires. Recently superconducting properties of both LaFeAsO$_{0.9}$F$_{0.1}$ and SmFeAsO$_{0.7}$F$_{0.3}$ wires were reported by our group.[18, 19] In fact, the 1111 phase REFeAsO$_{1-x}$F$_x$ superconductors are mechanically hard and brittle and are therefore not easy to draw into the desired wire geometry. In addition, REFeAsO$_{1-x}$F$_x$ superconductor is difficult to be synthesized because the sintering temperature is quite high (nearly 1200$^o$C) and the treatment time is long (over 50 hours). All of the above present the shortcomings of using such wires for applied purposes. However, the 122 phase $A$Fe$_2$As$_2$ superconductor seems much more suitable for making superconducting wires. First, the structure of $A$Fe$_2$As$_2$ superconductor is much simple compared to that of ZrCuSiAs-type ones. Second, this new type of $A$Fe$_2$As$_2$ superconductor can be easily synthesized due to very low heating temperature used (around 850 $^o$C), and several research groups have demonstrated the feasibility of fabricating large $A$Fe$_2$As$_2$ single crystals by self-flux method.[20, 21] So the 122 phase superconductor is a promising candidate for applications, however, there is no report about the fabrication of the $A$Fe$_2$As$_2$ superconducting wires so far. In this paper, we have fabricated niobium-clad $Sr_{0.6}K_{0.4}Fe_2As_2$ wires and investigated the superconducting properties of $Sr_{0.6}K_{0.4}Fe_2As_2$ wires.

The $Sr_{0.6}K_{0.4}Fe_2As_2$ composite wires were prepared with the in situ powder-in-tube (PIT) method using Sr, Fe, As and K as starting materials. The details of fabrication process are described elsewhere.[18] The raw materials were thoroughly grounded by hand with a mortar and pestle. The mixed powder was filled into an Nb tube of 8 mm outside diameter and 1 mm wall thickness. After packing, the tube was rotary swaged and then drawn to wires of 2.7 mm in diameter. The wires were cut into 4~6 cm and sealed in a Fe tube. They are then annealed at 850 °C for 35 hours. The high purity argon gas was allowed to flow into the furnace during the heat-treatment process to reduce the oxidation of the samples. It is noted that the grinding and packing processes were carried out in glove box in which high pure argon atmosphere is filled.

The phase identification and crystal structure investigation were carried out using x-ray diffraction (XRD) with Cu Kα radiation at 20-80° 2θ. The diffraction peaks could be well indexed on the basis of tetragonal $ThCr_2Si_2$-type structure with the space group I4/mmm, but some impurity phases are also detected, which were attributed to the unreacted material or unstable behavior of $Sr_{0.6}K_{0.4}Fe_2As_2$. The lattice constants were a=3.8882(6) Å and c=12.805(3) Å, consistent with the reported values.[10] Microstructure was studied using a scanning electron microscopy (SEM) after peeling away the Nb sheath. The superconducting properties of the wires were studied by magnetization and standard four-probe resistivity measurements using a physical property measurement system (PPMS). The critical current density $J_c$ was determined using the Bean model,[22] with the formulae $J_c = 20\Delta M/R$, where $\Delta M$ is the height of the magnetization loop and $R$ is the radius of the sample.

The transverse and longitudinal cross-sections of a Nb/ $Sr_{0.6}K_{0.4}Fe_2As_2$ wire after heat treatment are quite uniform and are shown in Figure 1. A reaction layer was observed between the superconducting core and Nb sheath. However, the thickness of reaction layer is much smaller than that of $SmFeAsO_{1-x}F_x$ wires due to the lower sintering temperature.[19, 23]

An expanded view of the resistive transition is plotted in figure 2 along with other curves measured in an applied field. At zero magnetic field we can observe a sharp transition with the onset temperature at 35.3 K and zero resistance at 29.5 K, which is close to 37 K measurement in the bulk superconductor.[10] Furthermore, applied magnetic field is observed to suppress the transition, as expected for superconducting transitions. It is noted that the onset transition temperature is not sensitive to magnetic field, but the zero resistance point shifts more quickly to lower temperatures due to the weak links or flux flow, which is similar to that of $AFe_2As_2$ single crystal.[5] The change of transition temperature ($T_c$) with critical field (H) is shown in the inset of Fig. 2, with $T_c$ defined at resistivity values corresponding to changes of 10 and 90 % in the total drop across $T_c$. It is clear that the curve of $H_{c2}(T)$ is very steep with a slope of $-dH_{c2}/dT|_{Tc} = 4.21$ T / K. From this Figure, using the Werthamer-Helfand-Hohenberg formula,[24] $H_{c2}(0) = 0.693 \times (dH_{c2}/dT) \times T_c$. we can get $H_{c2}(0) \approx 100$ T. If adopting a criterion of 99 % $\rho_n(T)$ instead of 90% $\rho_n(T)$, the $H_{c2}(0)$ value of this sample obtained by this equation is higher than 140 T. We can see that the irreversibility field in the inset of Fig. 2 is rather high comparing to that in $MgB_2$. These high values of $H_{c2}$ and $H_{irr}$ indicate that this new superconducting wire has an encouraging application in very high fields.

Figure 3 shows the temperature dependence of magnetic susceptibility of $Sr_{0.6}K_{0.4}Fe_2As_2$ wires. The measurements were carried out in a magnetic field of 10 Oe on heating after zero-field cooling and then on cooling in a field. The wire sample shows a well diamagnetic signal and superconductivity with $T_c = 32.7$ K, a superconducting volume fraction is large enough to constitute bulk superconductivity.

Figure 4 shows the SEM images illustrating the typical microstructure of the fractured core layers for $Sr_{0.6}K_{0.4}Fe_2As_2$ wires. It can be seen that the $Sr_{0.6}K_{0.4}Fe_2As_2$ wire seems to have a dense microstructure; however, some voids are formed by evaporation of Sr or As and the grains are not well connected. In other words, week links exist at the grains boundaries of this sample, and thus introduce a strong limitation to the flow of currents, so optimizing of processing parameters such as packing and deforming is needed to increase the density of $Sr_{0.6}K_{0.4}Fe_2As_2$ wires. The crystal structure of this kind of materials is based on a stack of alternating Sr(K) and FeAs layers. The grain size is about 10 μm. A layer-by-layer growth pattern can clearly be seen as shown in Fig. 4(b), very similar to what has been observed in Bi-based cuprates.

In Figure 5 we show the magnetization hysteresis loops (MHL) measured at different temperatures from 5 to 30 K for $Sr_{0.6}K_{0.4}Fe_2As_2$ wire samples. The gap in the loop drops drastically in the low-field region with increasing applied field but it rapidly reaches a slowly decreasing state over a wide range of applied magnetic field. The symmetric curves indicate that the bulk current instead of the surface shielding current dominates in the samples. It is remarkable that the superconducting MHL can still be measured at temperatures very close to $T_c$, with only a weak magnetic back ground. This suggests that the wire contains a small quantity of magnetic impurities, which is quite different from $SmFeAsO_{0.7}F_{0.3}$ wires.[19]

The critical current density $J_c$ calculated from the width of the hysteresis loops using the Bean model is shown in Figure 6. The critical current density $J_c$ values of $3.7 \times 10^3$ A/cm$^2$ at 5 K obtained for the $Sr_{0.6}K_{0.4}Fe_2As_2$ samples is significantly lower than that seen in random bulks of $MgB_2$ which generally attained $10^6$ A/cm$^2$ at 4.2 K, however, $J_c$ exhibits a rapid decrease at low-fields followed by a very weak field dependence at high-fields, showing that the $Sr_{0.6}K_{0.4}Fe_2As_2$ has a fairly large pinning force in the grain. It is noticed that $Sr_{0.6}K_{0.4}Fe_2As_2$ wire shows a very small magnetic field dependence $J_c$ even at temperatures very close to $T_c$.

The iron-based superconductor is the first non-copper-oxide superconductor with $T_c$ exceeding 50 K, which provides an excellent opportunity to be used in high fields. One crucial aspect of practical applications is the wire fabrication. Compared to 1111 phase $REFeAsO_{1-x}F_x$ wires, $Sr_{0.6}K_{0.4}Fe_2As_2$ wire is not brittle and is easy to draw into the desired wire geometry. Furthermore, the sintering temperature is far lower because of no oxygen, so $Sr_{0.6}K_{0.4}Fe_2As_2$ wire is much easier to be synthesized. The other important one for superconducting wire is the field dependence behavior especially at high fields. Although the critical temperature of $REFeAsO_{1-x}F_x$ superconductor is higher, our experimental data found that for $SmFeAsO_{1-x}F_x$ wires, the critical current $J_c$ decreases rapidly with the increase of the magnetic field, especially when the temperature is close to $T_c$, which is bad for applications in high fields. On this point, $Sr_{0.6}K_{0.4}Fe_2As_2$ wire is very advantageous; the $J_c$ exhibits a weak field dependence behavior even at high fields. Although the low temperature $J_c$ values are currently smaller than those of $MgB_2$, as of yet very few effort has been put into optimizing $J_c$. The processing parameters such as packing, deforming, treatment time and temperature have to be explored to increase the density and minimize the impurity phases. Applied directions of research will have to be explored in detail over the coming years.

In summary, we have fabricated Nd clad $Sr_{0.6}K_{0.4}Fe_2As_2$ wires by the PIT method. The $J_c$ of composite wires has a very weak dependence on magnetic field with $H_{c2}(0)$ above 140 T. Although some problems such as density and phase purity remain to be solved, our preliminary results explicitly demonstrate the feasibility of fabricating iron based superconducting wires. Of course, it is thus quite possible that we are still far away from the optimum processing conditions and the significant improvement can still be achieved.

The authors thank Profs. Haihu Wen, Liye Xiao and Liangzhen Lin for their help and useful discussion. This work was partly supported by the Natural Science Foundation of China (contract nos 50572104 and 50777062) and National '973' Program (grant no. 2006CB601004).

**Captions**

Figure 1 Optical images for a typical transverse (a) and longitudinal (b) cross-section of the $Sr_{0.6}K_{0.4}Fe_2As_2$ wires after heat treatment.

Figure 2 Resistivity at different fields of $Sr_{0.6}K_{0.4}Fe_2As_2$ wire. The inset shows the upper critical field line $H_{c2}$ and $H_{irr}$ as a function of the temperature. The $H_{c2}$ and $H_{irr}$ values were defined as the 90% and 10% points of the resistive transition, respectively.

Figure 3 Temperature dependence of magnetic susceptibility measured with H = 10 Oe for $Sr_{0.6}K_{0.4}Fe_2As_2$ wire.

Figure 4 (a) Low magnification and (b) high magnification SEM micrographs for the $Sr_{0.6}K_{0.4}Fe_2As_2$ wires.

Figure 5 Magnetization hysteretic loops of the $Sr_{0.6}K_{0.4}Fe_2As_2$ wires at different temperatures.

Figure 6 Magnetic field dependence of $J_c$ at different temperatures for $Sr_{0.6}K_{0.4}Fe_2As_2$ wires.

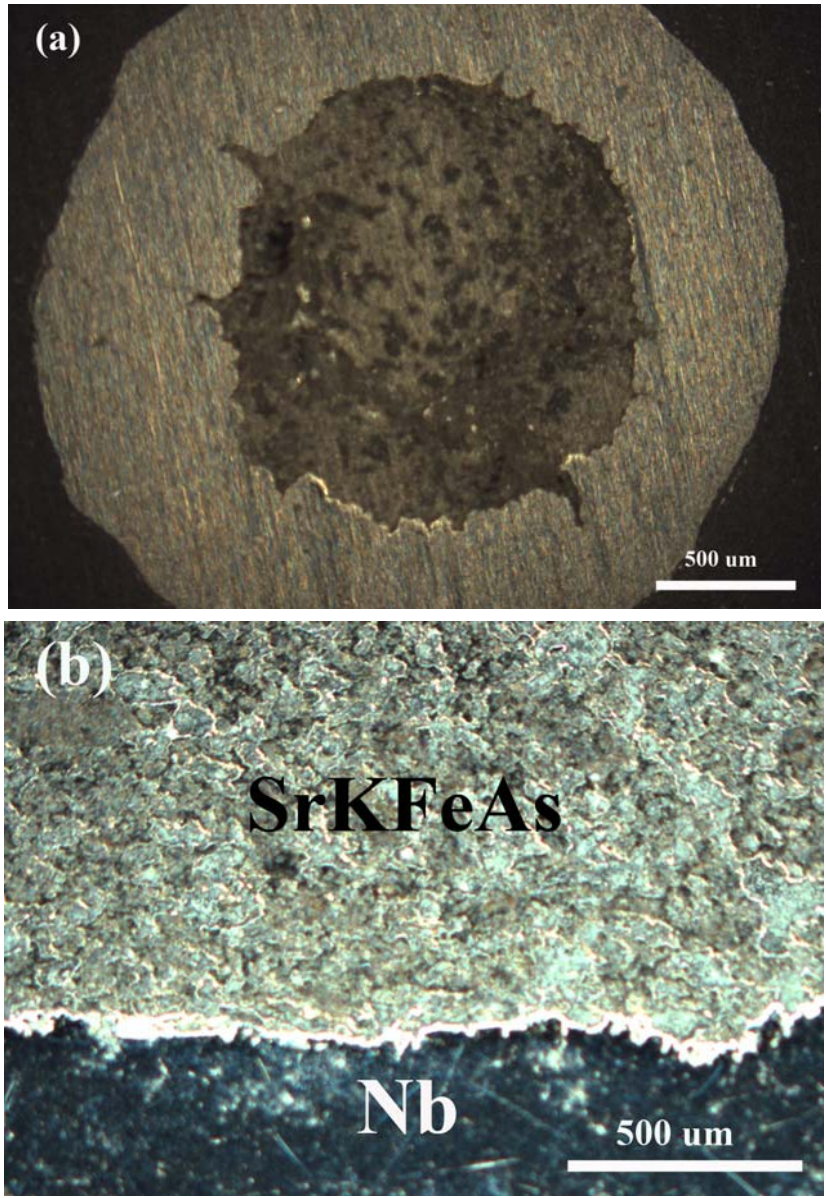

Fig.1 Qi et al.

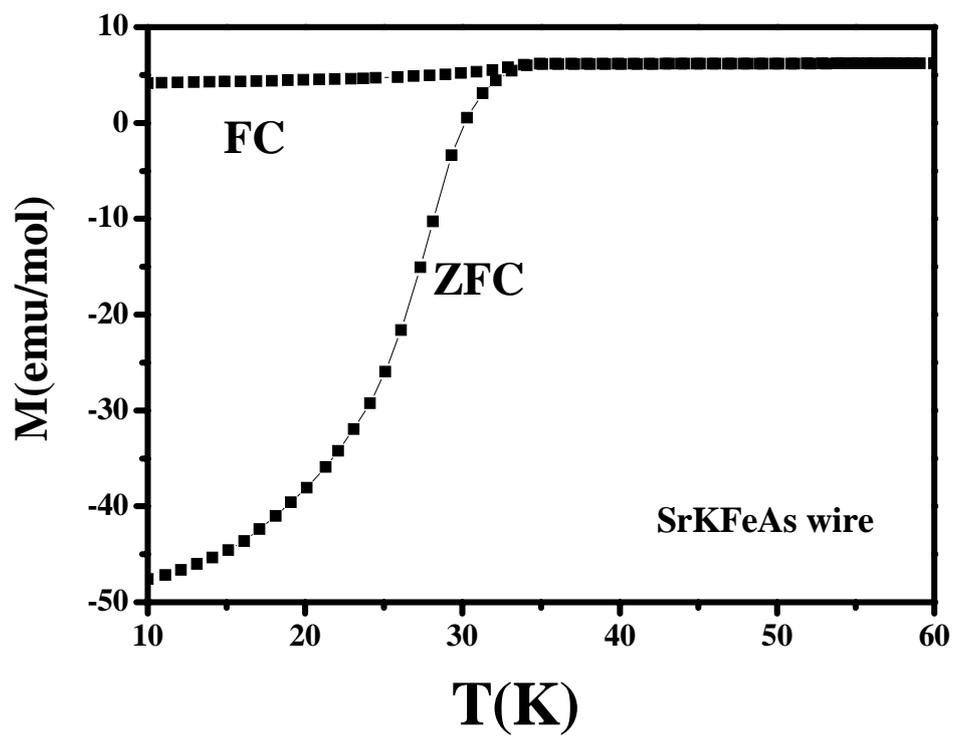

Fig.2 Qi et al.

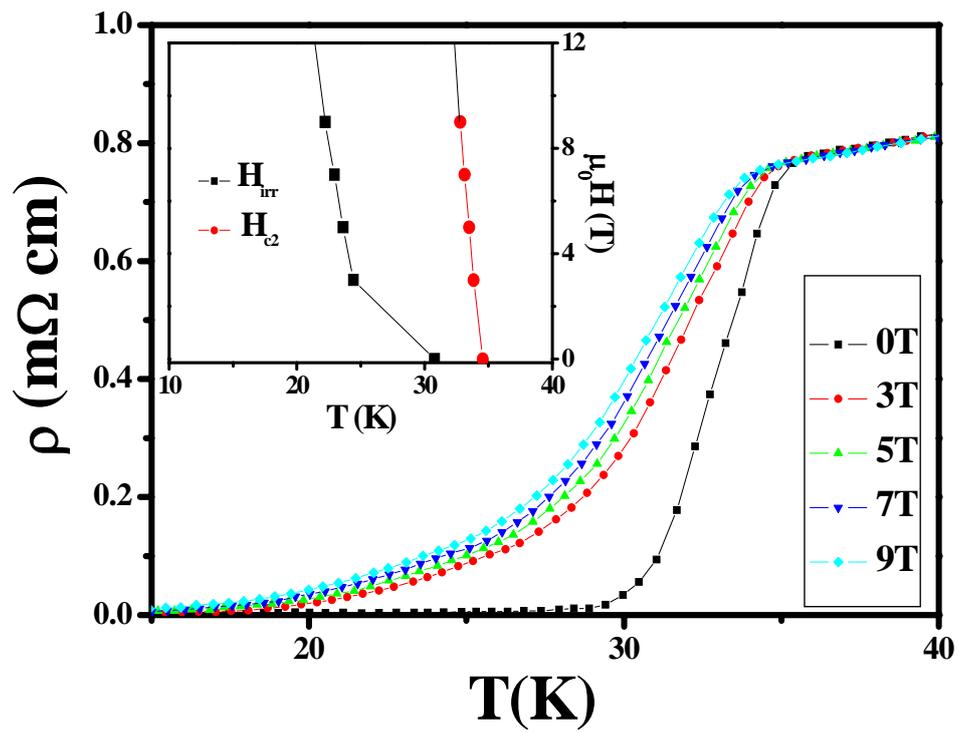

Fig.3 Qi et al.

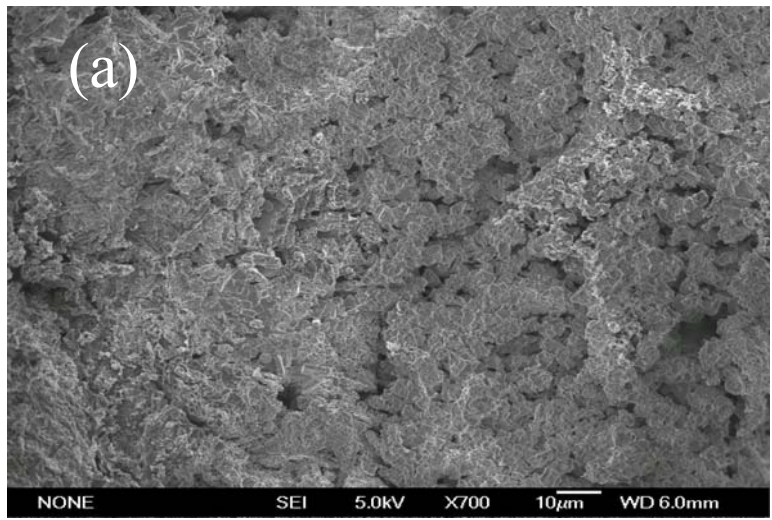
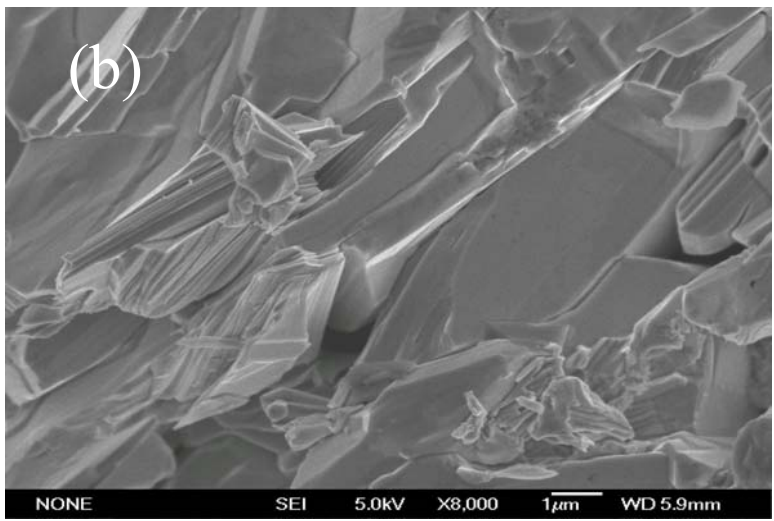

Fig.4 Qi et al.

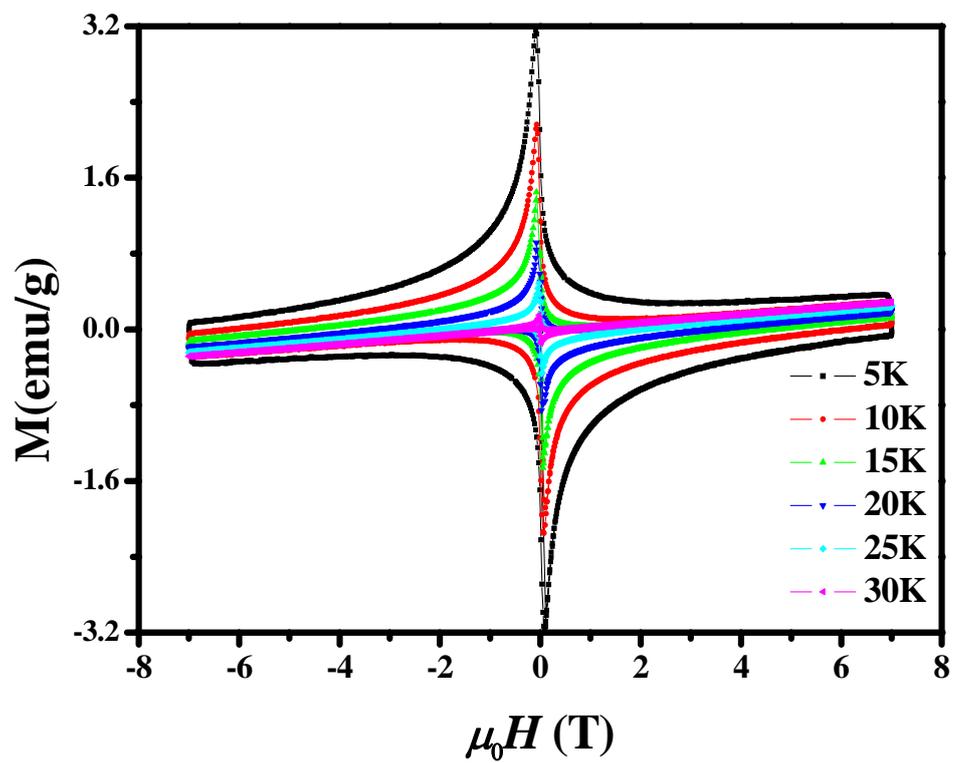

Fig.5 Qi et al.

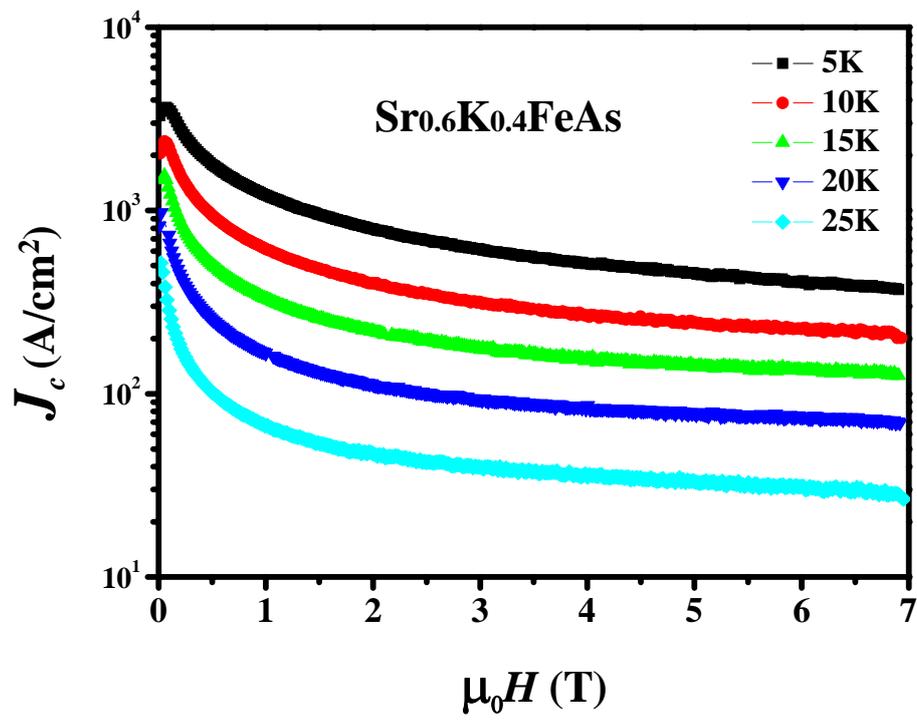

Fig.6 Qi et al.